\documentclass[conference, 9pt]{IEEEtran}
\IEEEoverridecommandlockouts

\usepackage{cite}
\usepackage{hyperref}
\usepackage{url} 
\usepackage{amsmath,amssymb,amsfonts}
\usepackage{algorithmic}
\usepackage{graphicx}
\usepackage{textcomp}
\usepackage{xcolor}
\usepackage{booktabs}
\usepackage{soul}
\usepackage{xcolor}
\usepackage{ulem}
\usepackage{xurl}
\makeatletter
\newcommand{\hlc}[2][yellow]{{%
  \colorlet{foo}{#1}%
  \sethlcolor{foo}\hl{#2}}%
}
\makeatother
\definecolor{oldlace}{HTML}{FFF4E4}
\definecolor{vanilla}{HTML}{F7E8A4}
\definecolor{mistryrose}{HTML}{FFEBE7}
\definecolor{tearose}{HTML}{E9BCB7}
\definecolor{uranianblue}{HTML}{ABDAFC}
\definecolor{cambriggeblue}{HTML}{8FC0A9}
\definecolor{azure}{HTML}{E5FCFF}
\definecolor{teacolor}{HTML}{E2F5D7}
\definecolor{oldrose}{HTML}{F2D4D5}
\definecolor{frenchgray}{HTML}{E7E9EC}

\def\BibTeX{{\rm B\kern-.05em{\sc i\kern-.025em b}\kern-.08em
    T\kern-.1667em\lower.7ex\hbox{E}\kern-.125emX}}
\begin{document}





\title{Can Large Audio-Language Models Truly Hear? Tackling Hallucinations with Multi-Task Assessment and Stepwise Audio Reasoning}


\author{\IEEEauthorblockN{1\textsuperscript{st} Chun-Yi Kuan}
\IEEEauthorblockA{\textit{Graduate Institute of Communication Engineering} \\
\textit{National Taiwan University}\\
Taipei, Taiwan \\
chunyi.kuan.tw@gmail.com}
\and
\IEEEauthorblockN{2\textsuperscript{nd} Hung-yi Lee}
\IEEEauthorblockA{\textit{Graduate Institute of Communication Engineering} \\
\textit{National Taiwan University}\\
Taipei, Taiwan \\
hungyilee@ntu.edu.tw}


}

\maketitle

\begin{abstract}
Recent advancements in large audio-language models (LALMs) have shown impressive capabilities in understanding and reasoning about audio and speech information. 
However, these models still face challenges, including hallucinating non-existent sound events, misidentifying the order of sound events, and incorrectly attributing
sound sources, which undermine their reliability and real-world application. 
To systematically evaluate these issues, we propose three distinct tasks: object existence, temporal order, and object attribute within audio. 
These tasks assess the models' comprehension of critical audio information aspects. 
Our experimental results reveal limitations in these fundamental tasks, underscoring the need for better models in recognizing specific sound events, determining event sequences, and identifying sound sources. 
To improve performance in these areas, we introduce a multi-turn chain-of-thought approach, which demonstrates significantly improved model performance across the proposed tasks.

\end{abstract}

\begin{IEEEkeywords}
Large audio-language models, hallucination
\end{IEEEkeywords}
\begin{figure*}[h] 
    \vspace{-10pt}
    \centering 
    \resizebox{0.80\textwidth}{!}{
    \includegraphics[width=\linewidth]{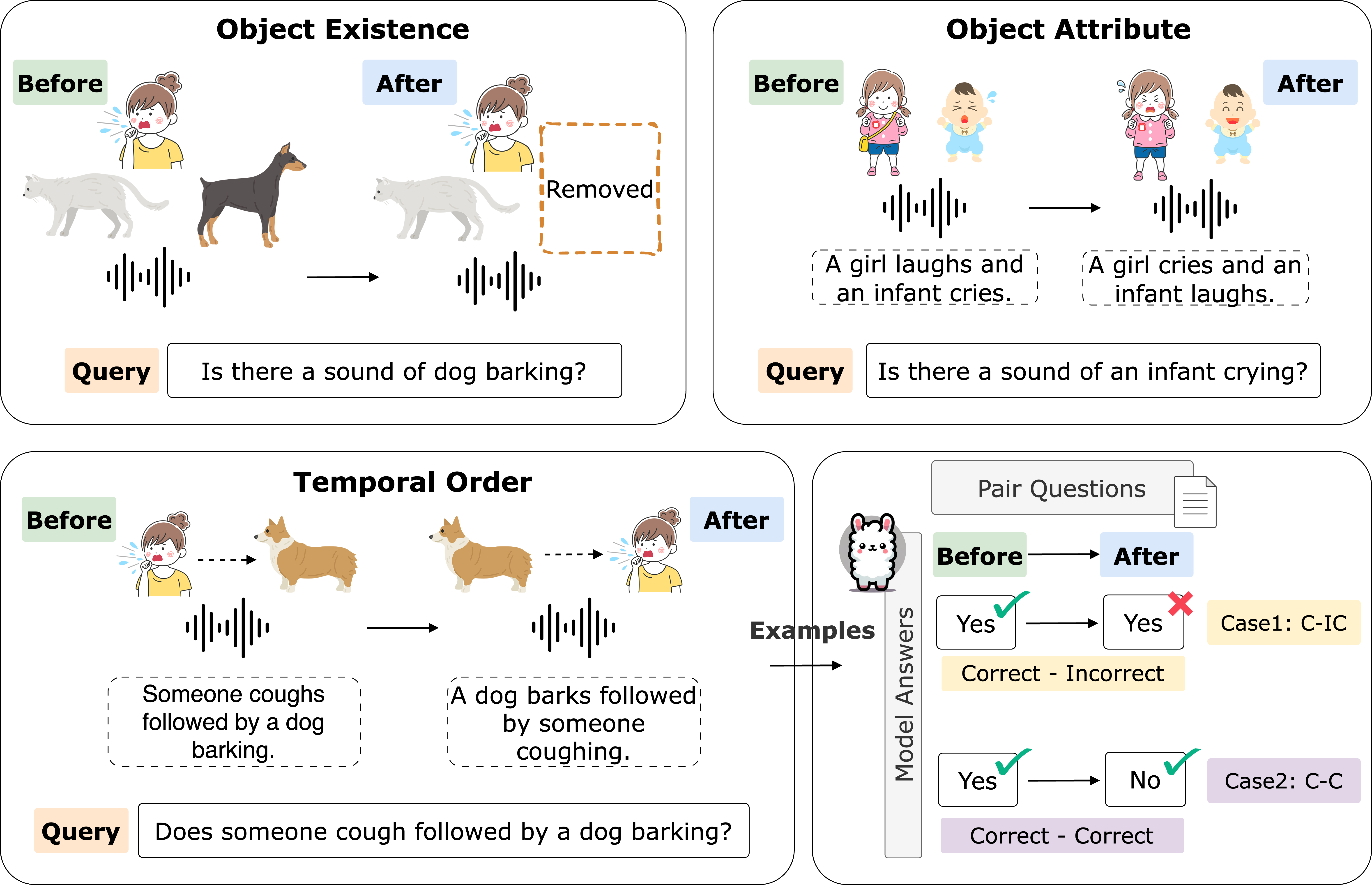} 
    }
    \caption{Demonstration of our evaluation pipelines.} 
    \label{fig:pipeline} 
\end{figure*}
\vspace{-5pt}

\section{Introduction}
 Large audio-language models (LALMs) \cite{chu2023qwen, chu2024qwen2, tang2023salmonn, gonglisten, gong_ltuas, hu2024wavllm, ghosh2024gama, kong2024audio, chen2023lauragpt, liu2023music, wu2023decoder, deshmukh2024pengi, wang2023slm, pan2023cosmic, fathullah-etal-2024-audiochatllama, wang2023blsp, wang2024blsp, kuan2024speech, lu2024desta} extend the capabilities of traditional text-based large language models (LLMs) by incorporating audio perception. 
These models can handle audio, speech, and text inputs simultaneously, using text prompts to process and extract relevant information from audio and speech.
These advancements in LALMs open up new possibilities for multimodal understanding and interaction. 
However, as with text-based language models\cite{rawte2023survey, zhang2023siren, huang2023survey, friel2023chainpoll, xu2024hallucination, ji2023survey, alkaissi2023artificial} and vision-language models\cite{wang2023evaluation, zhou2023analyzing, dai2023plausible, zhai2023halle, wang2023llm, liu2024survey}, the potential for hallucinations in LALMs raises important questions about their reliability and practical applications.

Investigating how these models experience hallucinations with audio inputs is crucial for several reasons.
First, existing benchmarks\cite{huang2024dynamic, yang2024air, weck2024muchomusic} primarily focus on instruction-based audio and speech tasks, lacking comprehensive evaluation of potential errors or hallucinations in audio understanding. 
Second, while some studies\cite{kuan2024understanding} have examined object hallucinations, they have not thoroughly explored the various aspects of hallucinations in audio. 
Finally, current research insufficiently addresses the need for models to not only recognize sounds but also understand their sequence and source, which are critical abilities for practical applications.
    
    

This research investigates model hallucinations in audio perception through a comprehensive analysis, focusing on sound identification, temporal order understanding, and sound source attribution. 
These aspects are crucial for assessing a model's fundamental audio comprehension abilities. 
For example, a model misidentifying a siren as music could lead to incorrect interpretations in emergency response systems. 
Similarly, a model might incorrectly identify a car horn as a trumpet, leading to misinterpretation of traffic scenarios in autonomous driving applications.
In critical applications such as security monitoring or medical diagnostics, sound recognition errors could have serious implications. 
For instance, misidentifying emergency signals could compromise safety. 

To explore model hallucinations, we develop tasks addressing object existence, temporal order, and object attributes, as illustrated in Fig. \ref{fig:pipeline}. 
For more examples, code, and benchmarks, visit our project page\footnotemark\label{footnote:lalm-hallucination}\footnotetext{ \url{github.com/kuan2jiu99/audio-hallucination}}.
Our dataset consists of 10,800, 3,116, and 1,614 instances for object existence, temporal order, and object attributes, respectively.
The object existence task evaluates the models' ability to detect specific sound events. 
The temporal order task assesses the models' capability to determine the sequence of sound events. 
The object attribute task examines the models' skill in identifying sound sources. 
We formulated these tasks as discriminative questions\cite{li2023evaluating}, such as asking about the presence of a dog barking or the relative timing of tiger and human sounds. 
These binary classification questions were evaluated using standard metrics such as accuracy, precision, recall, and F1 score.
To extend this approach, we introduced paired question sets inspired by previous research\cite{ye2024beaf}. 
These pairs consist of questions about two closely related audio segments: a ``before'' segment and an ``after'' segment. The ``after'' segment is identical to the ``before'' segment, except for the deliberate removal of one sound event.
This design assesses the models' ability to detect changes in sound events across sequential audio clips. 
For instance, in object existence section of Fig \ref{fig:pipeline}, a ``before'' clip might contain dog barking, cat meowing, and someone coughing, while the corresponding ``after'' clip would only include cat meowing and someone coughing.
By comparing the models' responses to questions about both segments, we can evaluate its sensitivity to changes in audio content over time.

Our findings indicate that current large audio-language models perform inadequately on these tasks. Significant improvements are needed in their capacity to detect objects, determine event order, and identify sound origins in audio inputs.
To address these issues, we propose the \textbf{M}ulti-turn \textbf{A}nd \textbf{T}houghtful \textbf{C}hain of \textbf{H}earings (MATCH) method. 
MATCH prompts the model to generate an audio description prior to answering specific questions. 
Our results show that this approach significantly improves model performance across all tasks.
In conclusion, our study contributes to the field in three main ways:
\begin{itemize}
    \item {We introduce three new aspects for evaluating hallucinations in audio-language models.}
    \item {We develop advanced before-and-after question designs to measure the model's ability to detect changes in sound events in audio.}
    \item {We propose a multi-turn and thoughtful chain of hearings method to improve models' performance on our designed tasks.}
\end{itemize}

\section{Method}

\begin{table*}[t]
\centering
\caption{Experimental results in percentages (\%). 
The metrics include A (Accuracy), P (Precision), R (Recall), F1 (F1 score), Yes (the proportion of ``yes" responses from the model), and IF (instruction following rate). 
The results are shown for four different settings: Orig. (original questions), Emp. (emphasis setting), Neg. (negative questions setting), and MATCH (the proposed Multi-turn And Thoughtful Chain of Hearings setting). 
Underlined text indicates the best performance.
We highlight the original F1 score with a gray background. 
Other settings with better F1 scores are highlighted brightly.}
\resizebox{\textwidth}{!}{
\small
\setlength{\tabcolsep}{2pt} 
\begin{tabular}{l c c c c c c c c | c c c c c c | c c c c c c | c c c c c c c}
\toprule
\multicolumn{1}{c}{} & \multicolumn{8}{c}{\footnotesize{Orig.}} & \multicolumn{6}{c}{\footnotesize{Emp.}} & \multicolumn{6}{c}
{\footnotesize{Neg.}} & \multicolumn{6}{c}{\footnotesize{MATCH}} \\
\cmidrule(lr){2-9} \cmidrule(lr){10-15} 
\cmidrule(lr){16-21} \cmidrule(lr){22-27}
\multicolumn{1}{c}{\textbf{Model}} 
& \textbf{A} & \textbf{P} & \textbf{R} & \textbf{F1} & \textbf{C-C} & \textbf{C-I} & \textbf{Yes} & \textbf{IF} 
& \textbf{A} & \textbf{P} & \textbf{R} & \textbf{F1} & \textbf{C-C} & \textbf{C-I} 
& \textbf{A} & \textbf{P} & \textbf{R} & \textbf{F1} & \textbf{C-C} & \textbf{C-I} 
& \textbf{A} & \textbf{P} & \textbf{R} & \textbf{F1} & \textbf{C-C} & \textbf{C-I} \\
\midrule
\multicolumn{1}{c}{\hlc[azure]{Object Existence}} \\
Qwen-Audio-Chat 
& 56.6 & 84.4 & 16.3 & \hlc[frenchgray]{27.3} & 5.2 & 92.1 & 90.3 & 100.0
& 59.7 & 79.7 & 26.2 & \hlc[azure]{39.4} & 9.6 & 83.9
& 54.9 & 81.8 & 13.2 & 22.7 & 4.3 & 93.0
& \underline{67.4} & \underline{74.5} & {53.2} & \hlc[azure]{62.1} & 22.0 & 58.2
\\
Qwen2-Audio-Instruct
& 54.4 & \underline{93.3} & 11.5 & \hlc[frenchgray]{20.5} & 3.8 & 95.4 & 92.2 & 98.3
& 61.1 & 85.2 & 34.3 & \hlc[azure]{48.9} & 14.4 & 79.3
& 54.6 & 77.6 & 16.9 & \hlc[azure]{27.8} & 5.1 & 89.9
& 67.1 & 71.0 & 58.0 & \hlc[azure]{63.8} & \underline{25.2} & 48.4
\\
SALMONN-7B 
& 53.9 & 76.9 & 11.3 & \hlc[frenchgray]{19.7} & 4.3 & 92.1 & 92.7 & 100.0
& 55.1 & 79.2 & 13.9 & \hlc[azure]{23.7} & 6.2 & 89.8
& 51.5 & 61.0 & 8.0 & 14.2 & 3.4 & 91.8
& 62.0 & 62.5 & 60.3 & \hlc[azure]{61.4} & 20.7 & 42.0
\\
SALMONN-13B 
& 59.5 & 84.0 & 23.5 & \hlc[frenchgray]{36.7} & 7.3 & 86.7 & 86.0 & 100.0
& 57.6 & \underline{88.3} & 17.4 & 29.1 & 4.8 & 92.0
& 53.8 & \underline{88.3} & 8.8 & 16.0 & 2.2 & 96.3
& 65.1 & 61.1 & 83.2 & \hlc[azure]{{\underline{70.5}}} & 22.0 & 22.7
\\
LTU-AS
& 53.1 & 54.2 & 39.7 & \hlc[frenchgray]{45.8} & 8.1 & 58.2 & 63.3 & 100.0
& 52.6 & 54.1 & 34.2 & 41.9 & 7.2 & 64.0
& 51.9 & 51.1 & 84.1 & \hlc[azure]{63.6} & 5.5 & 14.9
& 52.3 & 51.9 & 76.2 & \hlc[azure]{61.7} & 7.5 & 21.9
\\
Gazelle
& 50.2 & 50.3 & 66.7 & \hlc[frenchgray]{57.4} & 6.6 & 27.5 & 33.6 & 99.6
& 51.8 & 53.0 & 37.5 & 43.9 & 6.7 & 59.9
& 42.5 & 49.9 & 69.7 & \hlc[azure]{58.2} & 6.2 & 24.2
& 48.5 & 50.3 & 83.6 & \hlc[azure]{62.8} & 12.3 & 5.5
\\
GAMA
& \underline{62.2} & 68.9 & 44.7 & \hlc[frenchgray]{54.2} & 10.1 & 69.5 & 67.5 & 99.9
& \underline{64.2} & 66.8 & 56.6 & \hlc[azure]{61.3} & \underline{15.3} & 56.2
& \underline{60.3} & 62.3 & 76.9 & \hlc[azure]{68.8} & \underline{16.9} & 36.7
& 63.5 & 59.4 & 85.6 & \hlc[azure]{70.1} & 20.4 & 19.6
\\
Gemini-1.5-flash 
& 54.1 & 53.6 & 65.4 & \hlc[frenchgray]{58.9} & 10.2 & 33.6 & 38.9 & 99.2
& 53.0 & 52.1 & 77.2 & \hlc[azure]{62.2} & 9.1 & 20.8
& 42.2 & 51.8 & 83.2 & \hlc[azure]{63.8} & 6.6 & 17.1
& 45.3 & 50.8 & 97.1 & \hlc[azure]{66.7} & 2.2 & 1.6
\\
Gemini-1.5-pro 
& 46.0 & 51.6 & 96.6 & \hlc[frenchgray]{67.2} & 2.2 & 3.1 & 4.0 & 88.5
& 45.1 & 51.0 & \underline{98.4} & 67.2 & 1.6 & 1.4
& 41.2 & 51.1 & \underline{99.7} & \hlc[azure]{67.6} & 0.7 & 0.3
& 47.3 & 50.7 & \underline{97.9} & {66.8} & 2.3 & 1.2
\\
Cascade 
& 59.2 & 56.4 & \underline{97.6} & \hlc[frenchgray]{{\underline{71.5}}} & \underline{16.3} & 5.4 & 12.5 & 97.4
& 57.6 & 55.7 & 97.9 & \underline{71.0} & 13.9 & 4.8
& 57.4 & 56.8 & 96.8 & \hlc[azure]{{\underline{71.6}}} & 16.2 & 7.1
& - & - & - & - & - & -
\\
\midrule
\multicolumn{1}{c}{\hlc[teacolor]{Temporal Order}} \\
Qwen-Audio-Chat 
& 49.5 & 51.2 & 15.1 & \hlc[frenchgray]{23.3} & 9.3 & 75.4 & 84.0 & 98.5
& 48.8 & 49.8 & 38.1 & \hlc[teacolor]{43.2} & \underline{18.1} & 43.7
& 47.4 & 54.7 & 16.1 & \hlc[teacolor]{24.9} & 11.7 & 75.8
& 48.2 & 48.5 & 34.3 & \hlc[teacolor]{40.2} & 16.9 & 48.2
\\
Qwen2-Audio-Instruct
& 50.0 & {56.2} & 0.6 & \hlc[frenchgray]{1.1} & 0.4 & 99.2 & 99.4 & 99.9
& 50.6 & \underline{65.2} & 2.8 & \hlc[teacolor]{5.3} & 2.3 & 96.1
& 50.3 & {55.6} & 2.6 & \hlc[teacolor]{4.9} & 1.6 & 96.7
& 52.7 & {55.2} & 28.9 & \hlc[teacolor]{37.9} & 18.6 & 58.5
\\
SALMONN-7B
& 50.0 & 49.3 & 2.3 & \hlc[frenchgray]{4.4} & 1.7 & 96.2 & 97.7 & 100.0
& 50.6 & 58.2 & 4.1 & \hlc[teacolor]{7.7} & 2.9 & 94.4
& 49.9 & 37.5 & 0.2 & 0.4 & 0.3 & 99.4
& 52.5 & \underline{58.7} & 17.5 & \hlc[teacolor]{27.0} & 11.3 & 76.5
\\
SALMONN-13B
& \underline{53.4} & \underline{59.4} & 21.6 & \hlc[frenchgray]{31.6} & 12.6 & 74.6 & 81.8 & 100.0
& \underline{53.6} & 59.2 & 23.2 & \hlc[teacolor]{33.4} & 13.3 & 72.1
& \underline{50.4} & \underline{70.4} & 1.2 & 2.4 & 0.8 & 98.9
& \underline{58.1} & 55.2 & 85.8 & \hlc[teacolor]{{\underline{67.2}}} & \underline{23.9} & 8.5
\\ 
LTU-AS
& 50.5 & 50.6 & 44.7 & \hlc[frenchgray]{47.5} & 16.7 & 37.2 & 55.8 & 99.9
& 50.4 & 50.3 & 55.3 & \hlc[teacolor]{52.7} & 15.7 & 27.1
& 49.4 & 49.2 & 39.3 & \hlc[teacolor]{43.7} & 15.9 & 41.9
& 51.3 & 51.8 & 43.2 & \hlc[teacolor]{47.1} & 19.0 & 40.1
\\
Gazelle
& 47.9 & 49.7 & 32.8 & \hlc[frenchgray]{39.5} & 9.9 & 58.4 & 64.7 & 96.8
& 49.0 & 49.5 & 30.4 & 37.7 & 11.2 & 56.9
& \underline{50.4} & 50.5 & 39.0 & \hlc[teacolor]{44.0} & \underline{17.3} & 46.8
& 48.7 & 50.6 & 83.6 & \hlc[teacolor]{63.1} & 13.2 & 6.3
\\
GAMA
& 49.9 & 40.0 & 0.1 & \hlc[frenchgray]{0.3} & 0.1 & 99.7 & 99.7 & 99.8
& 50.1 & 59.3 & 1.0 & \hlc[teacolor]{2.0} & 0.9 & 98.3
& 49.3 & 40.0 & 1.4 & \hlc[teacolor]{2.7} & 0.8 & 97.1
& 50.3 & 50.7 & 24.9 & \hlc[teacolor]{33.4} & 12.0 & 66.0
\\
Gemini-1.5-flash
& 43.5 & 49.2 & 24.2 & \hlc[frenchgray]{32.5} & 12.1 & 62.4 & 66.0 & 87.5
& 39.5 & 48.8 & 67.5 & \hlc[teacolor]{56.7} & 10.4 & 18.0
& 18.0 & 49.9 & 74.8 & \hlc[teacolor]{59.9} & 8.3 & 18.3
& 44.0 & 51.0 & \underline{95.6} & \hlc[teacolor]{{{66.5}}} & 11.6 & 0.4
\\
Gemini-1.5-pro
& 35.7 & 52.7 & 66.0 & \hlc[frenchgray]{58.6} & \underline{19.9} & 18.1 & 24.6 & 67.9
& 40.1 & 50.7 & 86.0 & \hlc[teacolor]{63.8} & 9.8 & 5.5
& 26.0 & 50.5 & \underline{94.7} & \hlc[teacolor]{{\underline{65.9}}} & 3.7 & 1.5
& 44.9 & 51.5 & 94.3 & \hlc[teacolor]{66.6} & 10.7 & 0.7
\\
Cascade
& 40.6 & 50.3 & \underline{91.7} & \hlc[frenchgray]{{\underline{65.0}}} & 5.5 & 2.5 & 6.3 & 81.3
& 37.7 & 51.3 & \underline{89.9} & \hlc[teacolor]{{\underline{65.4}}} & 5.0 & 3.3
& 31.5 & 51.7 & 87.2 & \hlc[teacolor]{{{64.9}}} & 5.8 & 5.4
& - & - & - & - & - & -
\\

\midrule
\multicolumn{1}{c}{\hlc[oldrose]{Object Attribute}} \\
Qwen-Audio-Chat
& 50.0 & 54.2 & 1.6 & \hlc[frenchgray]{3.1} & 1.2 & 97.4 & 98.4 & 99.9
& 49.0 & 46.0 & 3.6 & \hlc[oldrose]{6.7} & 3.2 & 92.6
& 50.1 & \underline{59.4} & 2.4 & \hlc[oldrose]{4.6} & 1.9 & 96.5
& 51.2 & \underline{55.1} & 16.2 & \hlc[oldrose]{25.1} & 11.7 & 75.1
\\
Qwen2-Audio-Instruct
& 49.2 & \underline{61.4} & 3.5 & \hlc[frenchgray]{6.5} & 2.7 & 95.1 & 94.4 & 97.1
& 47.1 & \underline{57.4} & 7.8 & \hlc[oldrose]{13.8} & 5.5 & 88.6
& 48.1 & 49.0 & 9.6 & \hlc[oldrose]{16.0} & 4.5 & 85.4
& \underline{51.6} & 52.0 & 42.9 & \hlc[oldrose]{47.0} & \underline{17.5} & 42.9
\\
SALMONN-7B
& 49.8 & 46.0 & 2.9 & \hlc[frenchgray]{5.4} & 2.5 & 94.2 & 96.9 & 100.0
& \underline{50.3} & 55.1 & 3.4 & \hlc[oldrose]{6.3} & 2.7 & 94.6
& \underline{50.4} & 52.4 & 9.7 & \hlc[oldrose]{16.4} & 6.8 & 84.4
& 50.2 & 50.5 & 18.7 & \hlc[oldrose]{27.3} & 9.7 & 72.0
\\
SALMONN-13B
& 50.0 & 50.0 & 2.4 & \hlc[frenchgray]{4.5} & 2.2 & 95.4 & 97.7 & 100.0
& 49.9 & 48.6 & 2.1 & 4.0 & 1.9 & 95.9
& 50.1 & 54.6 & 1.5 & 2.9 & 1.5 & 97.3
& \underline{51.6} & 52.0 & 42.9 & \hlc[oldrose]{47.0} & \underline{17.5} & 42.9
\\
LTU-AS
& \underline{50.3} & 50.3 & 47.1 & \hlc[frenchgray]{48.6} & 13.6 & 39.8 & 53.2 & 100.0
& 48.8 & 48.6 & 42.8 & 45.5 & 12.9 & 41.9
& 50.0 & 50.0 & 83.8 & \hlc[oldrose]{62.6} & 7.3 & 8.9
& 50.7 & 50.6 & 65.7 & \hlc[oldrose]{57.1} & 13.4 & 22.3
\\
Gazelle
& 46.8 & 48.2 & 54.4 & \hlc[frenchgray]{51.1} & 13.6 & 28.2 & 42.5 & 97.1
& 47.7 & 48.6 & 59.7 & \hlc[oldrose]{53.6} & 12.5 & 24.7
& 43.1 & 50.3 & 63.5 & \hlc[oldrose]{56.1} & 15.3 & 23.4
& 50.7 & 50.6 & 72.1 & \hlc[oldrose]{59.5} & 8.4 & 21.6
\\
GAMA
& 49.1 & 43.2 & 6.0 & \hlc[frenchgray]{10.5} & 3.8 & 88.4 & 93.1 & 100.0
& 49.4 & 47.0 & 8.8 & \hlc[oldrose]{14.9} & 5.7 & 84.4
& 43.3 & 49.1 & 27.1 & \hlc[oldrose]{34.9} & 11.5 & 60.4
& 51.4 & 51.7 & 42.8 & \hlc[oldrose]{46.8} & 17.2 & 42.8
\\
Gemini-1.5-flash
& 49.3 & 50.6 & 48.2 & \hlc[frenchgray]{49.4} & \underline{18.6} & 34.2 & 51.1 & 97.5
& 49.2 & 50.2 & 58.5 & \hlc[oldrose]{54.1} & 16.7 & 25.3
& 41.6 & 49.7 & 64.5 & \hlc[oldrose]{56.1} & 14.8 & 20.6
& 42.3 & 51.0 & 79.0 & \hlc[oldrose]{62.0} & 11.2 & 9.5
\\
Gemini-1.5-pro
& 41.8 & 49.8 & \underline{86.1} & \hlc[frenchgray]{{\underline{63.1}}} & 9.9 & 3.1 & 11.8 & 83.5
& 40.7 & 49.7 & \underline{89.4} & \hlc[oldrose]{{\underline{63.9}}} & 7.2 & 2.9
& 37.9 & 49.9 & \underline{94.8} & \hlc[oldrose]{{\underline{65.4}}} & 3.7 & 0.4
& 42.0 & 50.2 & \underline{87.4} & \hlc[oldrose]{{\underline{63.8}}} & 7.5 & 4.1
\\
Cascade
& 48.5 & 50.6 & 61.8 & \hlc[frenchgray]{55.7} & 16.3 & 22.9 & 37.1 & 95.7
& 49.0 & 52.1 & 63.1 & \hlc[oldrose]{57.1} & 17.9 & 22.8
& 47.7 & 50.4 & 58.5 & 54.2 & 17.7 & 26.2
& - & - & - & - & - & -
\\

\bottomrule
\end{tabular}
}
\vspace{-10pt}
\label{tab:result}
\end{table*}

\subsection{Object Existence}
The object existence task evaluates the models' ability to detect specific sound events. 
We source data from the AudioCaps\cite{kim2019audiocaps}, ESC-50\cite{piczak2015dataset}, and VocalSound\cite{gong2022vocalsound} datasets. 
We use the test-split portion of AudioCaps, which contains numerous audio clips and their captions. 
These audio clips are treated as background sounds, denoted as \(B\).
The ESC-50 dataset includes 5-second-long recordings organized into 50 semantic classes, which can be divided into five categories: Animals, Natural soundscapes \& water sounds, Human non-speech sounds, Interior/domestic sounds, and Exterior/urban noises. The VocalSound dataset test-split includes recordings of laughter, sighs, coughs, throat clearing, sneezes, and sniffs. We use the data from ESC-50 and VocalSound as sound events.

Our process is as follows: First, select a sample from AudioCaps, referred to as the background audio \( B \). Second, extract three different sound events from ESC-50 or VocalSound, denoted as \( S_A \), \( S_B \), and \( S_C \), which are different from the background audio \( B \). Third, normalize all audio files to ensure a consistent volume range. Finally, mix the background audio \( B \) with sound event \( S_A \), \( S_B \), and \( S_C \) in sequence using an overlay method.
This process yields two paired audio samples:
\[
Pair_1 = B + S_A + S_B
\]
\[
Pair_2 = Pair_1 + S_C = B + S_A + S_B + S_C
\]
Here, the ``+'' symbol indicates the overlay operation. This way, we obtain a pair of audio samples that differ by the presence or absence of a specific sound event \( S_C \).

Therefore, an audio sample may contain multiple sounds, such as \( S_A \), \( S_B \), and \( S_C \). 
We test the model’s ability to recognize specific sounds by asking questions like ``Is there a sound of \( S_C \) in the audio?'' 
At the same time, there will be another paired audio sample where one of the sounds, for example, \( S_C \), is removed. 
This second audio sample will then only contain \( S_A \) and \( S_B \).
We query the model about the sound that was removed (in this case, \( S_C \)) using a pair of questions based on the original and modified audio samples. 
This allows us to test whether the model can accurately understand the presence or absence of specific sounds.

\subsection{Temporal Order}
The temporal order task assesses the models' capability to determine the sequence of sound events. 
We source data from the CompA-order\cite{ghoshcompa} dataset. 
The original CompA-order dataset includes 400 test instances, where each instance contains at least two audio-caption pairs. 
These pairs feature the same sound events, but the order of the events is different. 
In this way, each audio sample contains two distinct sounds with a clear temporal relationship. 
We test the model's understanding of temporal information in the audio by using prompt text such as ``Does the sound of a dog barking occur before the sound of a car horn honking in the audio?''
At the same time, there is a paired audio sample that includes the same two sounds but with the opposite temporal order. 
Such paired sample is presented with the same prompt text.
This approach of using paired questions allows us to better evaluate whether the model truly comprehends the temporal relationships between sounds in the audio.
By presenting both of these samples with the same question, we can more effectively assess the models' ability to discern and accurately report on the order of sounds in audio clips.

\subsection{Object Attribute}
The object attribute task examines the models' skill in identifying sound sources. 
We utilize data from the CompA-attribute\cite{ghoshcompa} dataset. 
The original dataset comprises 200 test instances, each containing two audio-caption pairs. 
These pairs feature the same sound event but are produced by different sources. 
For instance, ``An infant cries when a woman laughs'' and ``A woman cries when an infant laughs'' form a pair. 
These paired data are also used to create before-after question sets.
Moving on to the object attribute task, each audio sample includes two entities producing two distinct sounds. 
In our case, we have an infant crying and a woman laughing. 
We assess the models' ability to accurately identify which entity produces which sound by using prompt texts such as ``Is there a sound of an infant crying in the audio?'' and ``Is there a sound of a woman crying in the audio?''
Concurrently, we employ a paired audio sample where the sound-entity relationships are reversed - in this scenario, a woman crying and an infant laughing. 
We then query the model using the same prompt texts for this paired sample.
This method of using paired samples with reversed sound-entity relationships enables us to more thoroughly evaluate whether the model truly comprehends the information in the audio. 
It helps us determine if the model can accurately attribute sounds to their correct sources, rather than making assumptions based on common associations.
Testing both samples with the same questions helps us check how well the model can match sounds to their sources, even when they're unexpected.

\vspace{-3pt}
\section{Experimental Setups}
\vspace{-3pt}
\subsection{Baselines}
\begin{itemize}
    \item For open-source large audio-language models, we select Qwen2-Audio-7B-Instruct\cite{chu2024qwen2}, Qwen-Audio-Chat-7B\cite{chu2023qwen}, SALMONN-7B\cite{tang2023salmonn}, SALMONN-13B\cite{tang2023salmonn}, LTU-AS\cite{gong_ltuas}, Gazelle\cite{Gazelle}, and GAMA\cite{ghosh2024gama}.
    For closed-source models, we select Gemini-1.5-flash-001\cite{reid2024gemini} and Gemini-1.5-pro-001\cite{reid2024gemini}.

    
    
    \item In addition to end-to-end models, we also include a cascade pipeline as a baseline for comparison. 
    The cascade pipeline involves first using a specialized audio captioning model to generate captions for the audio. 
    Then, the description of the audio is fed into text-based large language models, which use this information to answer the corresponding questions. 
    In this setup, we used EnCLAP\cite{kim2024enclap} as the specialized audio captioning model, and LLaMA-3.1-8B-Instruct\cite{dubey2024llama} as the large language models. 
    In all experiments, all models use greedy decoding without any system prompt settings.

    \item As our questions are discriminative, we expect models to follow instructions and respond with ``yes'' or ``no''. 
    When parsing model responses, we use exact match to extract these answers. 
    If an answer could not be extracted, the result is excluded and reflected in the instruction following rate.
\end{itemize}

\vspace{-5pt}
\subsection{Ablation Study}
In addition to the original question format, we designed the following experimental settings:
\paragraph {Quotation Marks} Using quotation marks to emphasize the sound event in the question, for example, is there a ``cat meowing'' in the audio? We refer to this as the \textbf{emphasis} setting. 
In addition, we also use bold text to emphasize the sound event in the question, for instance, is there a \textbf{cat meowing} in the audio? 
\paragraph {Negative Questions} Reformulating the original question into a negative question, such as changing ``Is there a cat meowing in the audio?'' to ``Isn't there a cat meowing in the audio?''. 
We refer to this as \textbf{negative questions} setting.
\paragraph {Silent Input}The original audio input is replaced with silent input fed into the model.
\paragraph {Multi-turn and Thoughtful Chain of Hearings} 
Previous work \cite{kuan2024understanding} highlights that large audio-language models are proficient at audio captioning but not as effective in answering discriminative questions. 
Thus, our approach first involves having the model describe the audio information before prompting it to answer the question. 
The steps are outlined as follows: 
In the first round of interaction, the model is asked to generate captions, denoted as \({Cap}_{{audio}} \), for the given audio. 
In the second round, based on the previous dialogue history, the model is then asked to answer a corresponding question, \( {Q}_{{orig}} \).
We call this approach \textbf{MATCH}, which stands for \textit{\textbf{M}ulti-turn \textbf{A}nd \textbf{T}houghtful \textbf{C}hain of \textbf{H}earings}.
For instance, in the first round, the prompt "Describe the audio" is used to generate a caption for the audio, resulting in \({Cap}_{{audio}} \). 
At this point, the first round dialogue \texttt{\(\text{Dialogue}_{\text{first}} \) = <user> Describe the audio. <assistant> \(\text{Cap}_{\text{audio}} \)}.
Next, in the second round, the model is queried with the original question \({Q}_{{orig}}\), and its response is \({Ans}_{{model}}\). 
Thus, the second round dialogue becomes: 
\texttt{\(\text{Dialogue}_{\text{second}} \) = <user> Describe the audio. <assistant> \(\text{Cap}_{\text{audio}} \) <user> \(\text{Q}_{\text{orig}}\) <assistant> \(\text{Ans}_{\text{model}}\)}. 
For instance, consider an audio clip of a busy intersection. 
In the first round, the model might be prompted with ``Describe the audio'', to which it responds with a caption: ``The audio contains the sound of cars honking, people talking, and traffic signals beeping.'' 
This forms the first dialogue. 
In the second round, the model is asked a specific question like ``Is there a sound of thunder?'' 
Based on the previous caption and this question, the model might answer: ``No, there is no sound of thunder.''
For the temporal order task, we instead use the prompt: ``Describe the audio by focusing on the sequence and timing of sound events.''

\vspace{-8pt}
\section{Results}
In the original questions setting, open-source models demonstrate high precision but low recall and F1 score across the three task types, as shown in Table \ref{tab:result}. 
Both precision and recall metrics are based on hallucination questions, where the correct answer is ``No'', highlighting poor overall accuracy in handling questions with hallucination.
In contrast, commercial closed-source models have significantly higher recall, demonstrating better performance in answering hallucination questions. 
The cascade baseline achieve the best F1 score except for object attribute task. 
We also report the proportion of ``yes'' responses and instruction following rate (IF), noting that open-source models tend to predict ``yes'', while closed-source models and text-based LLMs exhibit the opposite trend.
Based on the performance on paired questions, baseline models struggle to consistently answer both correctly (Correct-Correct, C-C). 
This is also evident when models answer the ``before'' question correctly but fail on the ``after'' question (Correct-Incorrect, C-I). 
These results suggest that even if models can identify the content of the ``before'' segment, they often miss changes in the ``after'' segment, such as removing, reordering, or altering sound events.
We define C-C as the percentage of pairs where both ``before'' and ``after'' questions are correct, while C-I reflects cases where only the ``before'' question is correct. 
The experiments reveal challenges for LALMs beyond object hallucination, as noted in prior research \cite{kuan2024understanding}. 
Specifically, these models struggle with recognizing the correct temporal order and identifying sound sources in audio.
In the paired questions, although the text remains the same, the ``before'' segment's ground truth is ``yes'', while the ``after'' segment is ``no'' due to manipulations like removing or reordering sound events, or changing their sources. 
High C-I scores indicate that, even when models answer the ``before'' question correctly, they often fail to fully comprehend the audio. 
These findings underscore key areas where current models need improvement in audio understanding.

For the emphasis setting, some baseline models improve their recall and F1 scores when sound events were highlighted with parentheses. 
This is observed in models like Qwen-Audio-Chat, Qwen2-Audio-Instruct, and Gemini.
We also experiment with using bold text in markdown syntax to emphasize sound events, such as ``Is there a sound of **car horning**?", which also improves performance for some baselines. 
Due to space limitations, full results can be found here\footref{footnote:lalm-hallucination}.
In the negative questions setting, models like Qwen2-Audio-Instruct, GAMA, and Gemini show better recall and F1 scores with negative questioning. 
The above settings have little to no effect on the cascade pipeline. 

Our proposed method, MATCH, shows consistent improvements across all three tasks.
For the object existence task, open-source models demonstrate F1 score improvements ranging from approximately 10\% to 200\%.
The object attribute task shows improvements between 16\% and 944\%, where Gemini-1.5-flash also exhibits notable progress. 
In the temporal order task, most models show substantial improvements of 58\% or greater, with the exception of LTU-AS which exhibits only marginal enhancement. 
Notably, Gemini-1.5-flash also demonstrates performance gains in this task.
In most cases, the MATCH method simultaneously improves C-C performance and reduces C-I occurrences. 
This demonstrates its ability to enhance both accuracy and consistency in model outputs.
In conclusion, the MATCH setting consistently improves model performance across all tasks, particularly for object detection and temporal reasoning. 
These results demonstrate that generating audio descriptions before answering related questions significantly enhances model performance. 
This method effectively addresses the hallucination issues we identify. By using audio descriptions prior to question-answering, models are better equipped to handle tasks, reducing instances of hallucination and enhancing overall reliability in these tasks.

Additionally, in the silent setting, we replace the original audio with a silent signal and query the corresponding questions. 
We find that even with a silent signal, the models still exhibit severe hallucination effects. 
To explore this issue, we use the prompt ``Describe the audio.'' to query the models for audio captioning of the silent signal. 
Surprisingly, even with no sound present, the models would produce detailed descriptions, for example, ``The audio contains the sound of a car door closing nearby".  
It is note that the official Gemini documentation acknowledges limitations in non-speech sound recognition, stating that models supporting audio may make errors in this area. 
Moreover, Gemini sometimes responds with ``I'm just a text-based model and can't listen to sounds''. 
This limitation is reflected in its instruction following performance for audio tasks.

\vspace{-3pt}
\section{Conclusions}
\vspace{-1pt}
This study demonstrates that large audio-language models still struggle with hallucination across three distinct tasks: identifying specific sound events, distinguishing the order of sound events, and recognizing sound sources. 
To alleviate this issue, we introduce the \textbf{M}ulti-turn \textbf{A}nd \textbf{T}houghtful \textbf{C}hain of \textbf{H}earings (MATCH) method. 
By first asking models to provide audio-related descriptions before answering questions, MATCH significantly improves models performance on these tasks.

\section{Acknowledgement}
We would like to thank Wei-Ping Huang and Hsuan Su for providing valuable feedback on the draft of this paper. 
Additionally, we thank the National Center for High-performance Computing (NCHC) of National Applied Research 
Laboratories (NARLabs) in Taiwan for providing 
computational and storage resources.

\bibliographystyle{IEEEtran}
\bibliography{IEEEabrv, refs}

\end{document}